\documentclass{article}
\usepackage[utf8]{inputenc}
\usepackage{authblk}
\usepackage{setspace}
\usepackage[margin=1.25in]{geometry}
\usepackage{graphicx}
\graphicspath{ {./figures/} }
\usepackage{subcaption}
\usepackage{amsmath}
\usepackage{lineno}
\usepackage{multirow}
\usepackage{url}
\usepackage{hyperref}
\usepackage{xcolor}

\usepackage[style=ieee, 
citestyle=numeric-comp,
sorting=none]{biblatex}
\title{Determination of Double Beta Decay Half-life of $^{136}$Xe with the PandaX-4T Natural Xenon Detector}

\date{\today}

\begin{document}


\def\shKeyLab{School of Physics and Astronomy, Shanghai Jiao Tong University, MOE Key Laboratory for Particle Astrophysics and Cosmology, Shanghai Key Laboratory for Particle Physics and Cosmology, Shanghai 200240, China}
\def\BUAA{School of Physics, Beihang University, Beijing 100191, China}
\def\USTClab{State Key Laboratory of Particle Detection and Electronics, University of Science and Technology of China, Hefei 230026, China}
\def\USTCdep{Department of Modern Physics, University of Science and Technology of China, Hefei 230026, China}
\def\BUAALab{International Research Center for Nuclei and Particles in the Cosmos \& Beijing Key Laboratory of Advanced Nuclear Materials and Physics, Beihang University, Beijing 100191, China}
\def\pku{School of Physics, Peking University, Beijing 100871, China}
\def\YaLongSD{Yalong River Hydropower Development Company, Ltd., 288 Shuanglin Road, Chengdu 610051, China}
\def\IAP{Shanghai Institute of Applied Physics, Chinese Academy of Sciences, 201800 Shanghai, China}
\def\CHEPpku{Center for High Energy Physics, Peking University, Beijing 100871, China}
\def\SDUdep{Research Center for Particle Science and Technology, Institute of Frontier and Interdisciplinary Science, Shandong University, Qingdao 266237, Shandong, China}
\def\SDUlab{Key Laboratory of Particle Physics and Particle Irradiation of Ministry of Education, Shandong University, Qingdao 266237, Shandong, China}
\def\UMD{Department of Physics, University of Maryland, College Park, Maryland 20742, USA}
\def\TDLee{Tsung-Dao Lee Institute, Shanghai Jiao Tong University, Shanghai, 200240, China}
\def\MESJTU{School of Mechanical Engineering, Shanghai Jiao Tong University, Shanghai 200240, China}
\def\SYSUphy{School of Physics, Sun Yat-Sen University, Guangzhou 510275, China}
\def\SYSUnuc{Sino-French Institute of Nuclear Engineering and Technology, Sun Yat-Sen University, Zhuhai, 519082, China}
\def\NKU{School of Physics, Nankai University, Tianjin 300071, China}
\def\FDU{Key Laboratory of Nuclear Physics and Ion-beam Application (MOE), Institute of Modern Physics, Fudan University, Shanghai 200433, China}
\def\USST{School of Medical Instrument and Food Engineering, University of Shanghai for Science and Technology, Shanghai 200093, China}
\def\SJTUSC{Shanghai Jiao Tong University Sichuan Research Institute, Chengdu 610213, China}
\def\Princeton{Physics Department, Princeton University, Princeton, NJ 08544, USA}
\def\MIT{Department of Physics, Massachusetts Institute of Technology, Cambridge, MA 02139, USA}
\def\SARI{Shanghai Advanced Research Institute, Chinese Academy of Sciences, Shanghai 201210, China}
\def\SPEIT{SJTU Paris Elite Institute of Technology, Shanghai Jiao Tong University, Shanghai, 200240, China}



\author[1]{Lin Si}
\author[2]{Zhaokan Cheng}
\author[1]{Abdusalam Abdukerim}
\author[1]{Zihao Bo}
\author[1]{Wei Chen}
\author[1,3]{Xun Chen}
\author[4]{Yunhua Chen}
\author[5]{Chen Cheng}
\author[6,7]{Yunshan Cheng}
\author[8]{Xiangyi Cui}
\author[9]{Yingjie Fan}
\author[10]{Deqing Fang}
\author[10]{Changbo Fu}
\author[11]{Mengting Fu}
\author[12,13]{Lisheng Geng}
\author[1]{Karl Giboni}
\author[1]{Linhui Gu}
\author[4]{Xuyuan Guo}
\author[1,*]{Ke Han}
\author[1]{Changda He}
\author[4]{Jinrong He}
\author[1]{Di Huang}
\author[14]{Yanlin Huang}
\author[1]{Zhou Huang}
\author[3]{Ruquan Hou}
\author[15]{Xiangdong Ji}
\author[16]{Yonglin Ju}
\author[1]{Chenxiang Li}
\author[5]{Jiafu Li}
\author[4]{Mingchuan Li}
\author[16]{Shu Li}
\author[8]{Shuaijie Li}
\author[17,18]{Qing  Lin}
\author[1,8,3,$\dag$]{Jianglai Liu}
\author[6,7]{Xiaoying Lu}
\author[11]{Lingyin Luo}
\author[17,18]{Yunyang Luo}
\author[1]{Wenbo Ma}
\author[10]{Yugang Ma}
\author[11]{Yajun Mao}
\author[1,3]{Yue Meng}
\author[6,7]{Nasir Shaheed}
\author[1]{Xiaofeng Shang}
\author[1]{Xuyang Ning}
\author[4]{Ningchun Qi}
\author[1]{Zhicheng Qian}
\author[6,7]{Xiangxiang Ren}
\author[4]{Changsong Shang}
\author[12]{Guofang Shen}
\author[4]{Wenliang Sun}
\author[15]{Andi Tan}
\author[1,3]{Yi Tao}
\author[6,7]{Anqing Wang}
\author[6,7]{Meng Wang}
\author[10]{Qiuhong Wang}
\author[1,19,$\dag$$\dag$]{Shaobo Wang}
\author[11]{Siguang Wang}
\author[2,5]{Wei Wang}
\author[16]{Xiuli Wang}
\author[1,3,8]{Zhou Wang}
\author[2]{Yuehuan Wei}
\author[5]{Mengmeng Wu}
\author[1]{Weihao Wu}
\author[1]{Jingkai Xia}
\author[15]{Mengjiao Xiao}
\author[5,$\S$]{Xiang Xiao}
\author[8]{Pengwei Xie}
\author[1]{Binbin Yan}
\author[14]{Xiyu Yan}
\author[1]{Jijun Yang}
\author[1]{Yong Yang}
\author[9]{Chunxu Yu}
\author[6,7]{Jumin Yuan}
\author[1]{Ying Yuan}
\author[10]{Zhe Yuan}
\author[15]{Dan Zhang}
\author[1]{Minzhen Zhang}
\author[4]{Peng Zhang}
\author[1]{Shibo Zhang}
\author[5]{Shu Zhang}
\author[1]{Tao Zhang}
\author[1]{Li Zhao}
\author[14]{Qibin Zheng}
\author[4]{Jifang Zhou}
\author[1]{Ning Zhou}
\author[12]{Xiaopeng Zhou}
\author[4]{Yong Zhou}

\affil[1]{\shKeyLab}
\affil[2]{\SYSUnuc}
\affil[3]{\SJTUSC}
\affil[4]{\YaLongSD}
\affil[5]{\SYSUphy}
\affil[6]{\SDUdep}
\affil[7]{\SDUlab}
\affil[8]{\TDLee}
\affil[9]{\NKU}
\affil[10]{\FDU}
\affil[11]{\pku}
\affil[12]{\BUAA}
\affil[13]{\BUAALab}
\affil[14]{\USST}
\affil[15]{\UMD}
\affil[16]{\MESJTU}
\affil[17]{\USTClab}
\affil[18]{\USTCdep}
\affil[19]{\SPEIT\vspace{1em}}

\affil[ ]{\vspace{1em} (PandaX-4T Collaboration)}


\affil[*]{Corresponding Author: ke.han@sjtu.edu.cn}
\affil[$\dag$]{Spokesperson: jianglai.liu@sjtu.edu.cn}
\affil[$\dag$$\dag$]{Corresponding Author: shaobo.wang@sjtu.edu.cn}
\affil[$\S$]{Corresponding Author: xiaox93@mail.sysu.edu.cn}



\maketitle

\begin{abstract}
Precise measurement of two-neutrino double beta decay~(DBD) half-life is an important step for the searches of Majorana neutrinos with neutrinoless double beta decay.
We report the measurement of DBD half-life of $^{136}$Xe using the PandaX-4T dual-phase Time Projection Chamber~(TPC) with 3.7-tonne natural xenon and the first 94.9-day physics data release.
The background model in the fiducial volume is well constrained {\it in situ} by events in the outer active region.  
With a $^{136}$Xe exposure of 15.5\,kg-year, we establish the half-life as $2.27 \pm 0.03 (\textrm{stat.})\pm 0.10 (\textrm{syst.})\times 10^{21}$ years.
This is the first DBD half-life measurement with natural xenon and demonstrates the physics capability of a large-scale liquid xenon TPC in the field of rare event searches.
\end{abstract}

Two-neutrino double beta decay~(DBD) is a second-order weak process with an extremely long half-life~\cite{Goeppert-Mayer:1935uil}.
Neutrinoless double beta decay~(NLDBD), during which no neutrino is emitted, would be direct evidence of the Majorana nature of neutrinos and a clear violation of lepton number conservation~\cite{Furry:1939qr,Avignone:2007fu}. 
Therefore, the experimental search for NLDBD is an essential aspect of physics beyond the Standard Model (BSM)~\cite{Cremonesi:2013vla, Agostini:2017jim, Dolinski:2019nrj}.
Various experimental projects aim to search for NLDBD in different candidate isotopes.
Among them, enriched $^{136}$Xe has been used by EXO-200, KamLAND-ZEN, NEXT, etc~\cite{EXO-200:2019rkq,KamLAND-Zen:2016pfg,NEXT:2015wlq, Chen:2016qcd}.
PandaX-II has conducted the first search for NLDBD of $^{136}$Xe with a dark matter (DM) detector~\cite{PandaX-II:2019euf}.
XENON1T, another natural xenon-based TPC, has demonstrated excellent detector performance at the MeV range~\cite{XENON:2020iwh} and performed a search for NLDBD recently~\cite{Aprile:2022qou}.
The half-life of $^{136}$Xe DBD has been measured~\cite{Albert:2013gpz,KamLAND-Zen:2019imh,NEXT:2021dqj} and the most precise result of $2.165\pm 0.016(\rm{stat.})\pm0.059(\rm{syst.})\times 10^{21}$ years is given by EXO-200~\cite{Albert:2013gpz}. 

Precise measurement of DBD half-life is an important step for NLDBD searches.
For a finite detector resolution, DBD events contribute to the background of the NLDBD region of interest. 
Furthermore, a measurement of DBD half-life can help determine the nuclear matrix elements (NME) of the decay process and may shine light on the NME of NLDBD~\cite{Engel:2016xgb,Rodin:2003eb,Simkovic:2007vu}. 
More recently, there have been growing interests in exploring BSM physics with the electron spectrum of DBD (e.g.~\cite{Deppisch:2020mxv,Deppisch:2020sqh,Bolton:2020ncv,Agostini:2020cpz}). 

In this letter, we present a new measurement of $^{136}$Xe DBD half-life with large statistics and a broad energy spectral range with the dual-phase xenon Time Projection Chamber~(TPC) of PandaX-4T~\cite{PandaX-4T:2021bab}.
This represents the first DBD half-life measurement with natural xenon as the target. 
PandaX-4T has been originally optimized for signals in the low energy keV range for DM searches~\cite{PandaX:2018wtu}.
For the MeV energy range, we developed dedicated energy calibration procedures, reconstruction algorithms, and discrimination of single- and multiple-site events.
We systematically select the innermost cleanest part of the detector as our fiducial volume (FV) while utilizing the outer region to robustly characterize the background distributions {\it in situ}. 

PandaX-4T detector is a large dual-phase xenon TPC with approximately 3.7 tonnes of natural xenon in the active volume~\cite{PandaX-4T:2021bab}.
The active volume is enclosed by the field cage on the side, and the gate and cathode electrodes on the top and bottom, respectively.
The cross-section of the field cage is a polygon with 24 sides and its inscribed circle has a diameter of 1.185~m.
The liquid xenon (LXe) inside the field cage is 1.185~m tall.
The top and bottom PMT arrays have 169 and 199 three-inch Hamamatsu PMTs installed.
All PMTs are read out with custom-designed bases for high voltage biasing and signal readout~\cite{Yang:2021hnn}.
PMT signals are recorded by commercial 14-bit, 250~MHz sampling rate CAEN digitizers. 

An energetic event in the active volume of the detector generates a prompt scintillation signal ($S1$) and a delayed electroluminescence signal ($S2$).
Energy deposition in the active volume generates ionization electrons and detectable prompt scintillation light signals $S1$.
The ionization electrons drift up to the top of the active volume where they are extracted into the gaseous region.
The electric field in the gaseous region is much stronger, causing electrons to undergo an electroluminescence process to generate $S2$.
The energy and position of each event can be calculated from the amplitude and temporal information of $S1$ and $S2$.
The vertical (denoted as Z) position of an event can be determined from the drift velocity of electrons and the time delay between $S1$ and $S2$.
The relative amount of photons seen by each PMT on the top  PMT arrays is used to reconstruct the positions in the horizontal X-Y plane.

In this analysis, we used the first data release of PandaX-4T, taken from Nov.\,28, 2020 to Apr.\,16, 2021.
The total exposure is 94.9 days and divided into five different data sets, as detailed in Ref.~\cite{PandaX-4T:2021bab}.
This analysis emphasizes the energy range from 440 to 2800~keV, where the detector response is calibrated with external high-energy gamma sources including $^{137}$Cs, $^{60}$Co, and $^{232}$Th.
The gamma sources are encapsulated in small, centimeter-scale stainless steel cylinders, and then inserted into dedicated calibration pipes near the detector. 
Due to the higher activities, $^{137}$Cs and $^{60}$Co are placed at 397~cm and 375~cm away from the central axis of the TPC respectively, while $^{232}$Th with lower activity is placed at 84~cm away. All of them are approximately at the same height as the geometrical center of the TPC active volume.

\begin{figure}[b!]
    \centering
    \includegraphics[width=0.7\columnwidth]{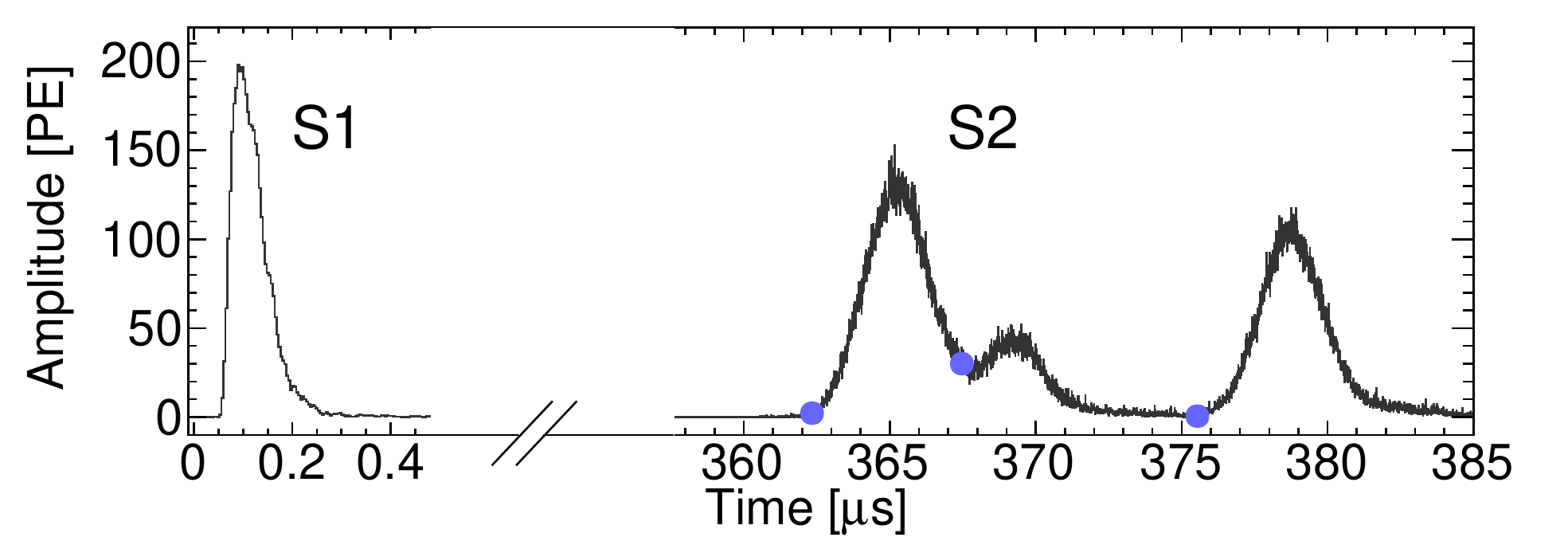}
    \includegraphics[width=0.7\columnwidth]{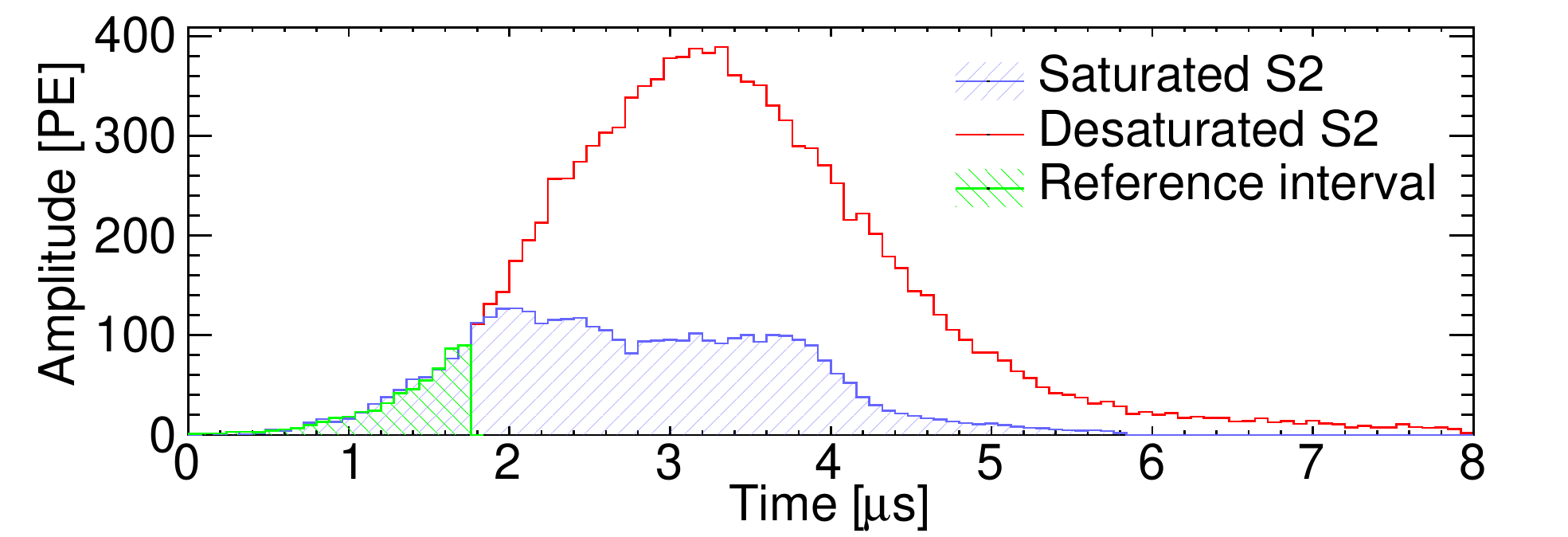}
    \caption{Top: Waveform of a sample MS event with one $S1$ peak and three $S2$ peaks identified by the MS algorithm. 
    The separation is indicated by the blue dots;
    Bottom: An example of saturated $S2$  waveform (blue) with a charge of approximately 3.8$\times$10$^3$~PEs. 
    The desaturated $S2$ (red) from a waveform template is scaled by matching the rising slope in the reference interval (green), resulting in a corrected charge of  $\sim 1.2\times10^4$~PEs.}
    \label{fig:pulses}
\end{figure}

For gamma events in the MeV range, energy loss via multiple Compton scattering dominates and can be identified via the timing profile of $S2$ signals.
In the summed $S2$ waveforms of PMT arrays, multiple site (MS) events register different pulses as shown in Fig.~\ref{fig:pulses} Top.
The $S2$ pulses of MS events can overlap in time due to the diffusion effect and misidentified as one single site (SS) event. 
To discriminate SS and MS events,  the summed pulse of all PMTs for a $S2$ is firstly smoothed by a locally weighted regression algorithm~\cite{Cleveland:1979}. 
We then identify peak(s) on the smoothed pulse, and a SS event is defined where all peaks are identified in a time window of $\pm0.75$~$\mu$s, corresponding to $\sim$2.2~mm in the vertical direction.
Otherwise, the events are labeled as MS events.

PMTs may suffer from the saturation effect with high energy $S2$ signals. 
For events in the MeV energy range, a typical PMT may collect up to thousands of photoelectrons (PEs).
We observe obvious waveform distortion when the total charge collected is above 900~PEs or so, as illustrated in Fig.~\ref{fig:pulses} Bottom.
PMT saturation effect is corrected by matching the rising slope of the saturated PMT waveforms to the non-saturated waveforms in the same event.
We use the summed PMT waveforms with charges in the range of 50 to 900~PEs as templates of non-saturated waveforms.
The waveform template is then scaled to match the rising slope of the saturated waveforms in a dynamic reference interval by minimizing the reduced $\chi^2$ between the two waveforms.
The scaled template offers an estimate of the true charge of saturated waveforms~\cite{XENON:2020iwh}.
We also validate the effectiveness of the desaturation with a bench measurement using PMTs illuminated by high-intensity photons with $S2$-like timing profiles.
The desaturation protocol is applied to the top PMTs for all events with distorted waveforms.
For events in the energy range of 1 to 3~MeV, the average correction factor is $\sim$3.0.
The S2 waveforms for the bottom PMTs are not corrected for saturation, but a residual correction is applied to the reconstructed energy (see later). 
On the other hand, we do not observe apparent saturation for $S1$ signals in the energy region of interest.

\begin{figure}[t!]
    \centering
    \includegraphics[width=0.7\columnwidth]{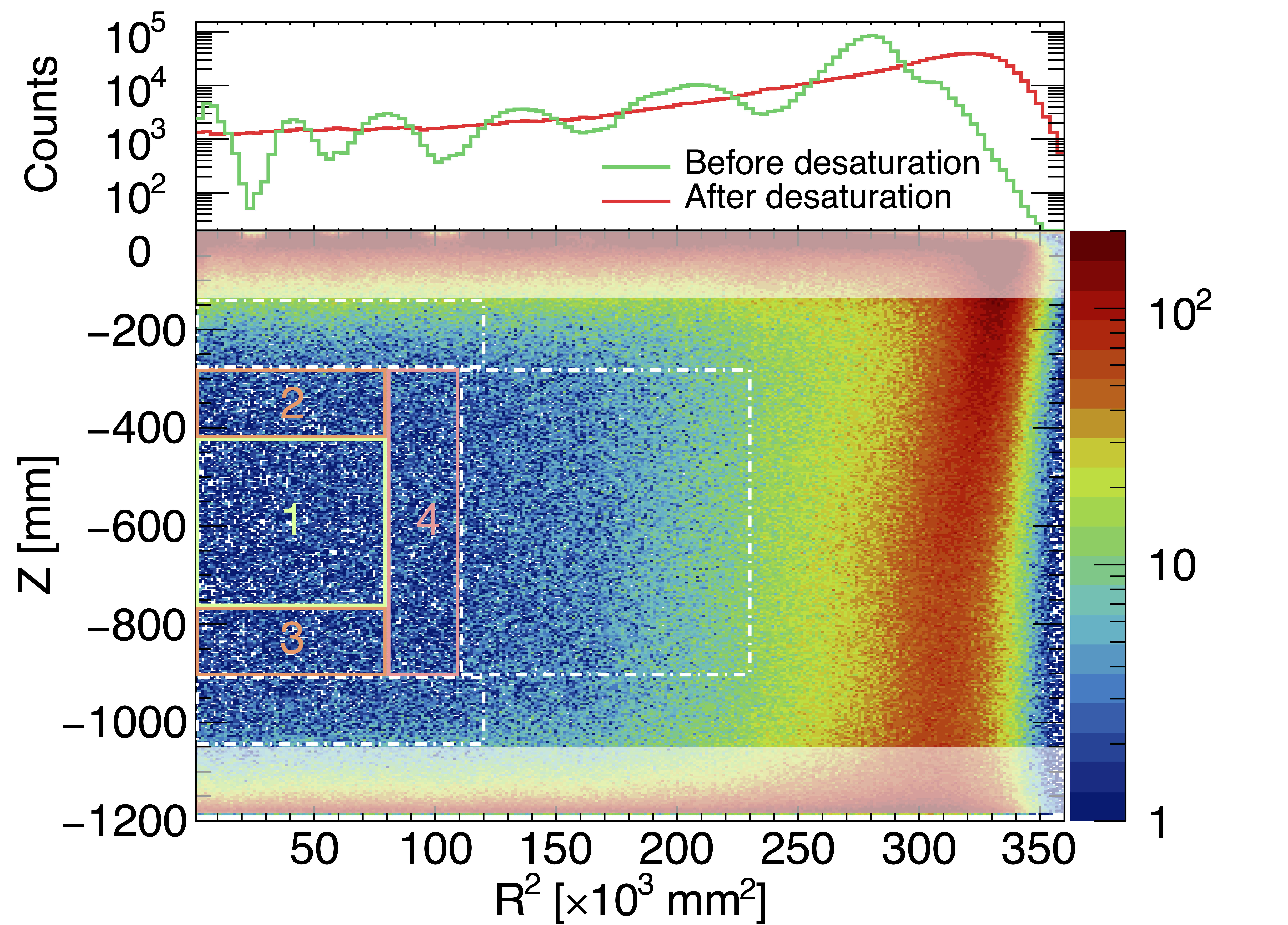}
    \caption{Top: Event position distribution in radius-squared (R$^2$) of the physics data in the energy range of [440, 2800]~keV reconstructed from the PAF algorithm with PMT charges before~(green) and after~(red) desaturation. 
    Bottom: Event distributions of the physics data in the energy range of [440, 2800]~keV in Z vs R$^2$ coordinates, with the color bar showing counts in each bin.
    The FV is divided into 4 regions, projected as 4 rectangles with numbers accordingly.
    The outer regions outlined by the dashed rectangles are used for cross-validation of the results obtained from the FV.
    The top and bottom shaded regions close to the PMT arrays are excluded in the top panel.}
    \label{fig:regions}
\end{figure}

Position reconstruction is significantly improved with desaturated PMT charges.
At high energy, we use the photon acceptance function (PAF) method~\cite{PANDA-X:2021jua} to reconstruct the position. 
PAF describes the expected charge detected by a certain PMT as a function of the event's position and is determined iteratively from data. 
A likelihood function is constructed to describe the charge distributions among each PMTs for an event, and the horizontal position is determined by maximizing the likelihood. 
In Fig.~\ref{fig:regions} Top, the distributions of positions reconstructed from desaturated and uncorrected PMT charges, respectively, are shown as a function of R$^2$, where R is the radial position.
The wiggles in the uncorrected curve are significantly smoothed out by the desaturation process.

The detector response to physical signals in different parts of the detector is non-uniform and the correction is done {\it in situ} with data.
$S2$ signals suffer from the absorption of electrons on electro-negative impurities in LXe while drifting up.
The electron lifetime is calculated with the 164~keV $\gamma$-ray peak from $^{131\rm m}$Xe run by run and used to correct $S2$ signals with an exponential function.
Electron loss may be convoluted with a saturation effect for large $S2$ signals, especially for the top PMT array.
Therefore, the electron lifetime correction is extracted based on $S2$ charges collected by the bottom PMT array (called $S2_b$), and the corresponding correction is applied to $S2_b$ in the physics data.
The average electron lifetime by dataset ranges from 800.4 to 1288.2~$\mu$s, while the maximum drift time is from 800 to 841~$\mu$s~\cite{PandaX-4T:2021bab}. 

Gaseous $^{83\rm m}$Kr calibration source is injected into the detector via the circulation loop~\cite{Zhang:2021shp} to map out the non-uniform response of the detector.
To get a response map to $S1$ signals, 41.5~keV events emitted from $^{83\rm m}$Kr are grouped into different voxels in the detector and the mean $S1$ charge in each voxel is calculated to represent the detector response at the corresponding positions. 
For the $S2$ signals, the response map is generated in the X-Y plane. 
The $S1$ and $S2$ signals for any event are corrected according to their position to achieve a uniform response throughout the active volume.

Energy calibration coefficients of SS events are inherited from the low energy analysis and applied with additional non-linearity correction.
Since the desaturation procedure introduces associated fluctuations, $S2_b$ without desaturation is used to reconstruct energy.
Although not being corrected with the desaturation algorithm directly, $S2_b$ benefits from improved position reconstruction and uniformity correction.
Energy is calculated from $S1$ and $S2_b$ according to the formula~\cite{Lenardo_2015} $E=13.7\,{\rm eV}\times\left(S1/{\rm PDE}+S2_b/({\rm EEE\times SEG_b})\right)$, where PDE, EEE, and $\rm{SEG_b}$ are the photon detection efficiency for $S1$, electron extraction efficiency, and the single-electron gain for $S2$ recorded by the bottom PMT array $S2_b$, respectively.
When applying the parameters for 5 datasets obtained from low energy analysis, we observe a 10\% level deviation for gamma peaks above 1 MeV, likely due to residual energy and position-dependent effects such as saturation.
As both $S1$ and $S2_b$ are not desaturated, $S1$ and $S2_b$ signals are further corrected to optimize the energy-dependent linearity for all locations in the detector.
The signals are corrected according to the reconstructed gamma peaks of the calibration data in different regions, and deviations of peak positions are reduced to the percent level.
The relative energy resolution $\sigma_E/E$ at the 2615\,keV (236\,keV) peak is 1.9\% (3.0\%) in the physics data and similar results are obtained at other peaks.
Considering the fact that the desaturation is not applied to  $S1$ and $S2_b$, an empirical function fit to the fitted energy resolution is used to smear the simulated spectra later to allow a data/simulation comparison.

Data quality selection cuts to remove unphysical events and select electron recoil events are developed based on calibration data, including that from $^{137}$Cs, $^{60}$Co, $^{232}$Th, $^{\rm{83m}}$Kr, and $^{220}$Rn.
Non-electron recoil and alpha events can be removed with a cut on the charge of $S1$ and/or $S2$ as well as the ratio between the two.
The relative size of $S1$ charges collected by the top and bottom PMT arrays and drift time can be cross-compared to reject accidental coincidence events and events originating from the gate electrode.
All cuts were optimized and determined with the last 9.6~days of data after going through the high-energy specific data processing, while the rest of the data is blinded.
The total cut efficiency is calculated to be $(99.4\pm0.4)\%$, while the uncertainty is estimated by the difference in different detector volumes.
These cuts are applied to all physics data, and the efficiency is validated to be the same as that from 9.6~days of data.

\begin{figure}[b!]
    \centering
    \includegraphics[width=0.7\columnwidth]{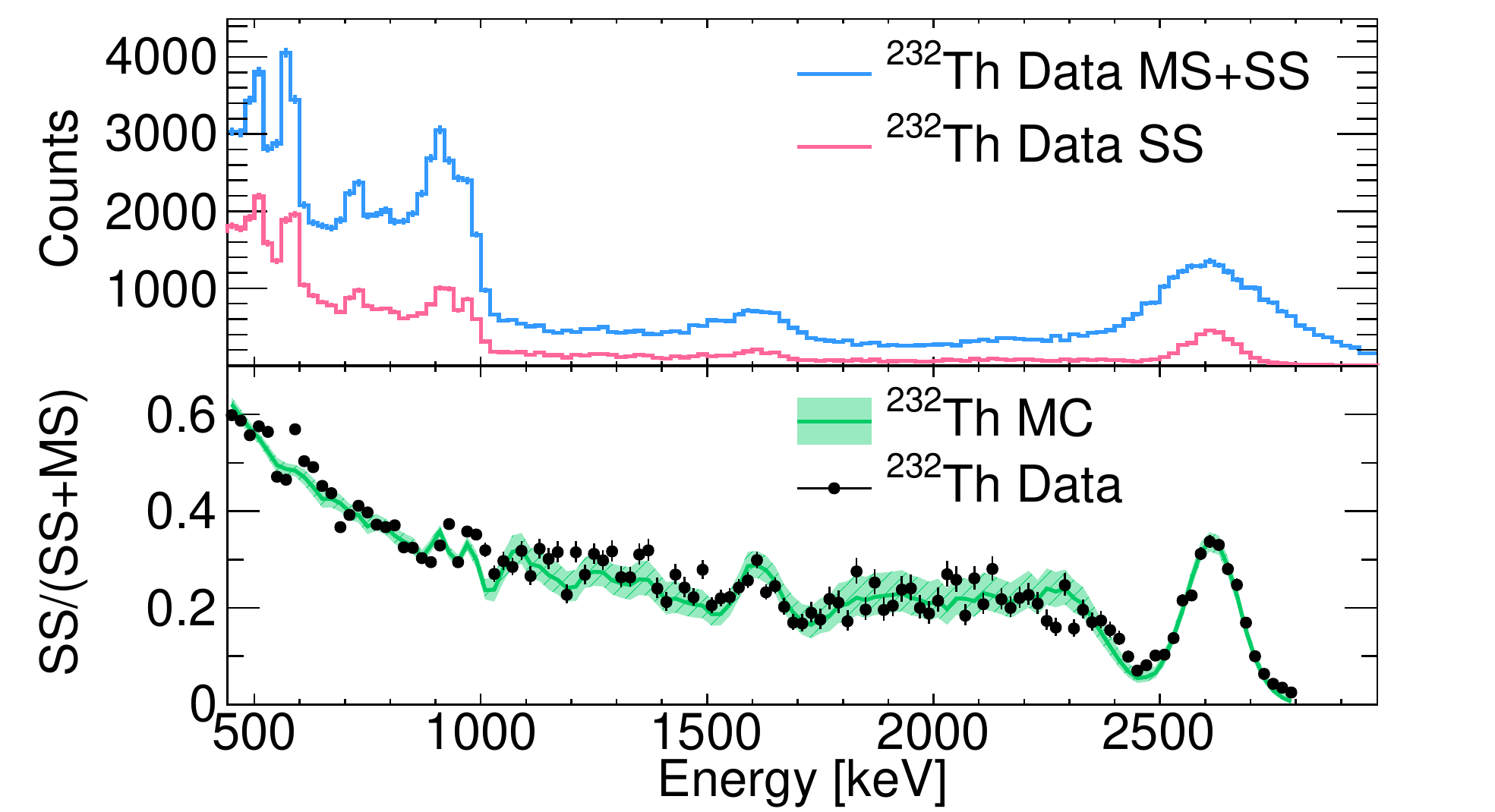}
    \caption{Top: SS (magenta) and MS+SS (cyan) spectra of $^{232}$Th calibration data. Bottom: Comparison of SS fraction between MC (shaded green) and data (black), with an average difference of 1.7\%.
    The uncertainties of the MC spectrum are shown in the green shaded band.}
    \label{fig:SSMS}
\end{figure}

Detailed simulation with Geant4-based Monte Carlo framework BambooMC~\cite{Chen:2021asx} is established with full detector geometry, including top and bottom PMT arrays with individual PMT and base, field cage, stainless steel vessels, supporting frame, and water shielding tanks.
LXe responses to ER in the MeV range are modeled from standard NEST 2.0 construction~\cite{Lenardo_2015,PandaX-II:2021jmq}, with light yield, charge yield, and recombination parameters extracted from calibration.
Energy depositions simulated in Geant4 are converted into individual $S2$s, which are then smeared in time with drift-time-correlated Gaussian diffusion measured in the data. 
The pseudo-$S2$ waveforms are then piped through the SS/MS discrimination algorithm.
The resulting SS and MS spectra are compared to the corresponding spectra in the calibration runs.
No systematic deviation is observed as a function of energy, and within 440~keV to 2800~keV, the overall agreement is at 1.7\% level (Fig.~\ref{fig:SSMS}).
When taking the bin-by-bin deviation into account, the fit result (see later) shifts by 1.5\%. Conservatively, we use 1.7\% as a systematic uncertainty in the final DBD half-life calculation.

The majority of the background events in our region of interest (ROI) from 440 to 2800\,keV are from radioactive contamination of detector components and impurities in the xenon.
Radioactive contamination of $^{60}$Co, $^{40}$K, and the U/Th chain from major detector components is assayed with dedicated setups such as high purity germanium detectors, with results summarized in Ref.~\cite{Qian:2021qiu}.
We generate expected background contributions in SS spectra according to simulation, assay results, and the discrimination algorithm. 
The major detector components are grouped into three categories, denoted as top, bottom, and side.
The top category includes the top flanges of the vessels and the top PMT assembly, which consists of the PMT array, readout circuits, cabling, and the mechanical supporting structure. 
The counterpart bottom PMT assembly and the bottom hemisphere of the vessels are grouped as the bottom category.
The side category is composed of the field cage and cylindrical barrel of the vessels. 
Other detector components are found to have negligible background contributions and thus not included. 
The weighted sums of expected background counts in the ROI from four radioactive contaminations are listed in Table~\ref{tb:components}.

$^{222}$Rn emanated from the inner surface of the detector and circulation pipes is the major internal contamination. 
$^{214}$Pb and $^{214}$Bi, progenies of the $^{222}$Rn, contribute mostly to the ROI of DBD.
97\% of the beta decays from $^{214}$Bi can be rejected together with their subsequent alpha decay of $^{214}$Po with a half-life of $\sim163~\mu$s~\cite{Zhu:2021qss}. 
The remaining $^{214}$Bi activity is less than  0.1~$\mu$Bq/kg, which makes a negligible contribution to our ROI.
Therefore, only the contribution from $^{214}$Pb is considered and simulated with BambooMC.
$^{85}$Kr beta decay also contributes to the lower end of the ROI with end-point energy of 687~keV.
However, with an extremely low concentration level at $^{85}$Kr/Xe ratio of $6.6\pm4.2\times 10^{-24}$~\cite{PandaX-4T:2021bab}, the tail of its beta spectrum has a marginal impact on our result and has not been included in the fit.

\begin{table}[b!]
    \centering
    \caption{Expected and fitted contribution of background contaminations originating from the top, bottom, and side of the detector and LXe inside. 
    All values are reported in the number of counts in the FV.}
    \label{tb:components}
    \begin{tabular}{cccc}
    \hline
    \hline
    Detector part & Contamination & Expected counts  & Fitted counts  \\ \hline
    \multirow{4}{*}{Top} & $^{238}$U & $339\pm129$ & $490\pm52$ \\
    & $^{232}$Th &  $402\pm133$ & $670\pm56$  \\
    & $^{60}$Co &  $327\pm141$ & $550\pm49$  \\
    & $^{40}$K &  $300\pm156$ & $363\pm40$  \\ \hline
    \multirow{4}{*}{Bottom} & $^{238}$U  & $141\pm51$  & $185 \pm40$ \\
    & $^{232}$Th &  $237\pm119$ & $155\pm53$ \\
    & $^{60}$Co &  $159\pm95$ & $183\pm48$  \\
    & $^{40}$K &  $89\pm84$ & $100\pm39$  \\ \hline
    \multirow{4}{*}{Side} & $^{238}$U  & $475\pm707$ & $1070\pm118$ \\
    & $^{232}$Th &  $786\pm959$ & $2194\pm117$  \\
    & $^{60}$Co &  $1244\pm945$ & $185\pm98$  \\
    & $^{40}$K &  $1518\pm835$ & $782\pm84$  \\  \hline
    LXe & $^{214}$Pb ($^{222}$Rn progeny) &  $[0, 12057]$ &   $7180\pm152$ \\
    \hline
    \hline
    \end{tabular}
\end{table}

DBD signal spectrum is also simulated with BambooMC.
The energy of two electrons from DBD is generated with the Decay0 package~\cite{Ponkratenko:2000um} as input for our simulation.
For DBD events with energy greater than 440~keV, the SS fraction is $97.4\%$ with a fractional uncertainty of 1.7\% according to our detector response simulation.

A cylindrical FV with a radius of 33.2\,cm and a height of 66.3\,cm in the geometrically center part of the detector is selected for the final fit, with the range in the Z direction pre-determined by the event rate distribution in the ROI.
The FV is then determined from $^{220}$Rn and $^{\rm{83m}}$Kr calibration data.
Both internal calibration sources are expected to be evenly distributed in the active volume.
The FV is defined where the proportionality between the event counts and geometrically calculated volume is better than 0.5\%, and the uncertainty of the FV cut is estimated as 1.0\%.
The FV is further divided into 4 regions, as shown in Fig.~\ref{fig:regions} Bottom and data spectra are reconstructed in each region.
Region 1 is the innermost and cleanest region.
Region 2, 3, and 4 are on outside of Region 1, where the external radioactive contaminations from the top, bottom, and side of the detector, respectively, have more impact.

We constructed a simultaneous fit with the binned likelihood function defined as
\begin{equation}
\begin{aligned}
    L = \displaystyle\prod_{i = 1}^{N_{\rm R}}\displaystyle\prod_{j = 1}^{N_{\rm{bins}}}\frac{(N_{ij})^{N_{ij}^{\rm{obs}}}}{N_{ij}^{\rm{obs}}!}e^{-N_{ij}}
    \displaystyle\prod_{k = 1}^{N_{\rm{bkgs}}}\frac{1}{\sqrt{2\pi}\sigma_{k}}e^{-\frac{1}{2}(\frac{\eta_{k}}{\sigma_{k}})^{2}}\,,
\label{eq:PL1}
\end{aligned}
\end{equation}
\begin{equation}
    N_{ij} = n_{\rm{Xe}}S_{ij}^{\rm{Xe}}+\displaystyle\sum_{k=1}^{N_{\rm{bkgs}}}(1+\eta_{k})n_kB_{ij}^{k}\,,
\label{eq:PL2}
\end{equation}

\noindent
where $N_{ij}^{\rm{obs}}$, $N_{ij}$ are the observed and expected event numbers in the $j$-th energy bin of the $i$-th region. 
$N_{ij}$ are modeled as Eq.~\ref{eq:PL2} according to PDFs of the DBD spectrum $S_{ij}^{\rm{Xe}}$ and 
the background components $B_{ij}^{k}$ given as (category, isotope) pairs and $^{214}$Pb listed in Table~\ref{tb:components}.
The PDFs are weighted by the number of counts $n_{\rm{Xe}}$ and $n_k$ for signal and backgrounds respectively.
For background, $n_k$ is fixed while the nuisance parameters $\eta_{k}$ denotes the fractional difference between the prior and fitted counts. 
In the likelihood function, a Gaussian penalty is added for each background except $^{214}$Pb based on the prior fractional uncertainty $\sigma_k$ derived from Table~\ref{tb:components}.
The prior $^{222}$Rn level is measured {\it in situ} with the number of high-energy alpha particles emitted in the decay chain with small uncertainty.
However, the number of $^{214}$Pb may be smaller due to progenies attaching to electrodes and the inner surface of the detector~\cite{Ma:2020kll}.
Therefore, the number of $^{214}$Pb events is unconstrained in the fit.

\begin{figure}[tbh]
    \centering
    \includegraphics[width=0.7\columnwidth]{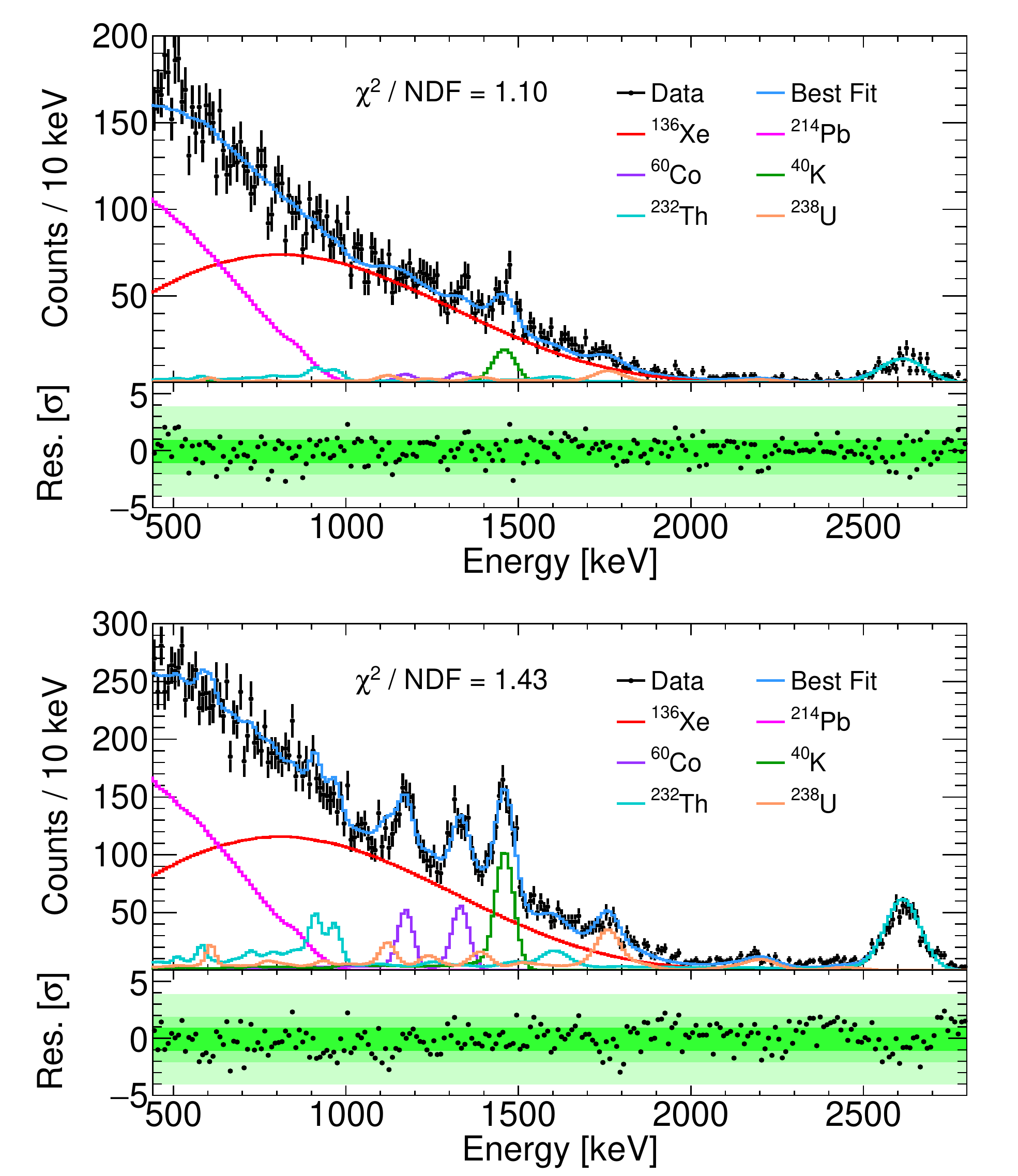}
    \caption{Results of the simultaneous fit. 
    For better visualization, Region 1 is shown individually (top) and the other three regions are shown together (bottom). 
    Fitted spectral of background components of (category, isotope) pairs are grouped and plotted based on isotopes.
    In each figure, data and fit results are shown in the top panel and residuals in the bottom panel.
    The $\chi^2$/NDF (number of degrees of freedom) as an indicator of the goodness of fit is shown for each figure.}
    \label{fig:fits}
\end{figure}

The fitted spectra in Region 1 and the combined results of the other three regions are shown in Fig.~\ref{fig:fits}. 
In the FV, $^{136}$Xe DBD events are dominant.
The total number of DBD events is $17468\pm 257$.
A parallel fit is also performed with the RooFit package~\cite{Verkerke:2003ir} giving consistent results.

The fitted results of all background sources are listed in Table~\ref{tb:components} in comparison with the expected counts. 
The most noticeable background is from $^{214}$Pb, the progeny of $^{222}$Rn.
The ratio of 59.6\% between the fitted count of $^{214}$Pb and the expected count of $^{222}$Rn, which is estimated from its alpha events, represents the aforementioned depletion of $^{214}$Pb.
The large sensitive volume of PandaX-4T helps determine the external radioactive contaminations more accurately and robustly, as demonstrated by the smaller uncertainties on the fitted number of counts.
Agreements within two sigmas between our fit results and radioactive assays are observed for most of the contributions.
For comparison, the best fit results are used to calculate the expected spectrum for the regions outside of the FV, as outlined by the three dashed rectangles in Fig.~\ref{fig:regions}.
Within our DBD ROI, the largest difference between expected and measured rates is 2.3\% and the agreement is within 1\% when the three dashed regions are considered together.

We performed the fit by varying bin size from 1~keV to 40~keV and lower (upper) fit range from 440 (2600) to 600 (3000)~keV.
The impact of both changes is at the 1\% level or smaller.
Systematic uncertainties may also come from the mismatch between simulated and measured spectra in the FV. 
Three independent effects are studied including uncertainties in energy resolution, energy scale, and the relative variation of weights among the four regions. 
To cleanly evaluate the systematic effect of energy reconstruction, the maximum deviation of both the reconstructed energy peak and energy resolution is included. 
PDFs from the simulation are generated with energy resolutions scaled from 0.8 to 1.2 times the measured resolutions. 
We also shifted the peak positions of PDFs by up to $\pm10$\,keV. 
The relative weight $n_{k}$ in Eq.~\ref{eq:PL2} is calculated from the geometry or scaled by the measured $^{222}$Rn rates in each region.
Different impact on DBD half-life between the two sets of weights is treated as the regional weight systematics, as listed in Table~\ref{tb:syst}. 
All the background PDFs are generated assuming secular equilibrium of the U/Th chains.
If the early and late parts of the decay chains are allowed to float independently in our spectrum fit, we observe a 2.0\% shift in our DBD result, mainly due to the thorium chain.
The shift is treated as a systematic uncertainty from the non-equilibrium decay chain.
Moreover, a recent re-evaluation of $^{214}$Pb beta decay spectrum pointed out that the spectrum calculated by Geant4 may not be precise~\cite{Haselschwardt:2020iey} 
In our ROI, the new theoretical calculation changes the spectrum by approximately -6\% to 30\% from low energy to high energy, if we assume the same correction can be applied to $^{214}$Pb decay to excited states as well as the ground state.
Therefore, a sizable impact of 2.0\% is observed in our fit result.
Similarly, $^{136}$Xe DBD shape may also influence the half-life measurement. 
An alternative shape assuming a single-state dominance hypothesis (SSD)~\cite{Kotila_SSD} is used in the spectrum fit and the result differs from the baseline choice by 0.36\%.

\begin{table}[tbp]
    \centering
    \caption{Summary of relative systematic uncertainties.} 
    \label{tb:syst}
    \begin{tabular}{cccc}
    \hline
    \hline
    Source~        & Percentage  & Source~  & Percentage  \\ \hline
    Quality cut~   & 0.39\%      & SS cut~  & 1.7\%      \\
    FV cut~        & 1.0\%      & Bin size~      & 0.05\%   \\
    Fit range~ & 1.2\%      & Energy resolution~  & 0.58\%     \\
    Energy scale~ & 0.26\%      & Regional weight~  & 1.6\%     \\
    $^{214}$Pb spectrum~ & 2.0\%     & LXe density~     & 0.13\%    \\
    $^{136}$Xe abundance~     & 1.9\%     & $^{136}$Xe spec. shape~ &0.36\% \\
    Non-equilibrium decay chain~    &2.0\%    & \textbf{Total} & \textbf{4.5\%} \\
    \hline
    \hline
    \end{tabular}
\end{table}

Aside from what has already been mentioned earlier, we have also evaluated the systematics related to the $^{136}$Xe abundance.
The fluctuation of gas phase pressure $2.064\pm0.008$~bar during the data taking leads to an LXe density of $2.8502\pm0.0036$\,g/cm$^3$, based on the saturated vapor properties of Xe.
Therefore, the total xenon mass in the FV is $649.7\pm6.5$\,kg, where the uncertainty comes from the FV cut and LXe density.
We adopt the reported $^{136}$Xe abundance of 8.86\% in natural xenon and measure the xenon samples from the PandaX-4T TPC by a residual gas analyzer (RGA)~\cite{RGA} for cross-check and error estimation.
With and without the relative ionization efficiencies of different xenon isotopes~\cite{NIST} taken into account, the measured isotopic abundance of $^{136}$Xe is ($9.03\pm0.03$)\% and ($8.79\pm0.03$)\%, respectively. Conservatively, we use ($8.86\pm0.17)\%$ in the half-life evaluation, resulting in a $^{136}$Xe mass of $59.6\pm1.3$\,kg in the FV.

\begin{figure}[tbh]
    \centering
    \includegraphics[width=0.7\columnwidth]{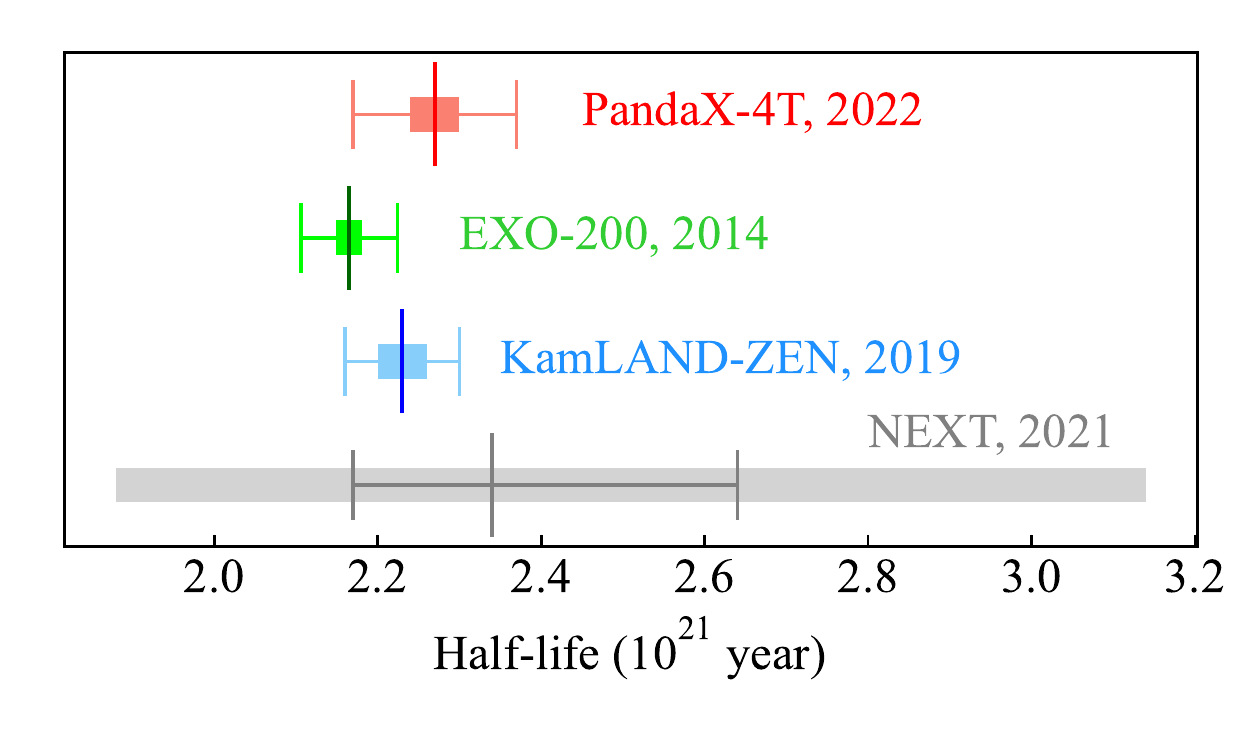}
    \caption{Comparison with half-life measurements from enriched xenon experiments~\cite{Albert:2013gpz,KamLAND-Zen:2019imh,NEXT:2021dqj}.
    The box and cap-tipped error bars represent statistical and systematic uncertainties respectively.}
    \label{fig:compare}
\end{figure}

The total relative systematic uncertainty is 4.5\%, including all the systematics listed in Table~\ref{tb:syst} summed in quadrature.
With a live time of 94.9 days, a selection and SS efficiency of 99.4\% and 97.4\%, respectively, and a DBD event fraction of 86.3\% within our fit range, we obtain a final DBD half-life measurement of $2.27 \pm 0.03 (\textrm{stat.})\pm 0.10 (\textrm{syst.})\times 10^{21}$ years.
This result agrees with the previous measurements from KamLAND-ZEN, EXO-200, and NEXT collaborations~\cite{Albert:2013gpz,KamLAND-Zen:2019imh,NEXT:2021dqj}, as shown in Fig.~\ref{fig:compare}.
Our measurement with PandaX-4T, which is initially designed as a dark matter direct detection experiment, demonstrates the physics reach of large liquid xenon TPC on multiple fronts.
Thanks to its emphasis on low energy, our analysis threshold is at least 260~keV lower compared to the previous measurements.
With a large active volume and self-shielding, we can aggressively select the innermost and cleanest part of the detector for the measurement. 
The outer part of the active volume is used to cross-check the background and signal model {\it in situ}. 
Such advantages can also be exploited in other physics searches in the MeV range in the future. 

In summary, we have measured the DBD half-life of $^{136}$Xe with the 94.9 days of PandaX-4T data. 
Our analysis uses an FV at the very center of the detector with $ 649.7\pm6.5$\,kg of xenon and a total number of $17468\pm 257$ DBD events was observed in the energy range of 440 to 2800~keV. 
The final result $2.27 \pm 0.03 (\textrm{stat.})\pm 0.10 (\textrm{syst.})\times 10^{21}$ years is one of the most precise measurements to date and the first result using a natural xenon dark matter detector with robust background control. 
Our DBD analysis is a major step forward for the NLDBD searches with a large-scale multiple-purpose liquid xenon TPC.
PandaX-4T has continued data taking with the physics goal of dark matter direct detection, double beta decay searches, and measurement of neutrinos from astrophysical sources. 
We are also developing upgrade plans to enhance the detector response at the MeV level and possibly lower the background rate by replacing detector components.
More and higher quality data will further improve our measurement of DBD half-life and understanding of the background relevant to NLDBD searches, and may be used to explore potential new physics~\cite{Deppisch:2020mxv,Deppisch:2020sqh,Bolton:2020ncv,Agostini:2020cpz}.

\section*{Data Availability}
The data used to support the findings of this study are available from the corresponding author upon request.

\section*{Conflicts of Interest}
The authors declare that there is no conflict of interest regarding the publication of this article.

\section*{Authors’ Contributions}
L. Si and Z.K. Cheng led the data analysis. 
K. Han, S.B. Wang, and X. Xiao drafted the manuscript. 
All authors on the list contributed to the construction of the PandaX-4T detector, as well as data acquisition and processing.

\section*{Acknowledgements}

 
This project is supported in part by grants from the Ministry of Science and Technology of China (No. 2016YFA0400301 and 2016YFA0400302), grants from National Science
Foundation of China (Nos. 12090061, 12090062, 12005131, 11905128, 11925502), 
and by Office of Science and
Technology, Shanghai Municipal Government (grant No. 18JC1410200). We thank supports from Double First Class Plan of
the Shanghai Jiao Tong University. We also thank the sponsorship from the
Chinese Academy of Sciences Center for Excellence in Particle
Physics (CCEPP), Hongwen Foundation in Hong Kong, and Tencent
Foundation in China. Finally, we thank the CJPL administration and
the Yalong River Hydropower Development Company Ltd. for
indispensable logistical support and other help.

\printbibliography

@article{Avignone:2007fu,
    author = "Avignone, III, Frank T. and Elliott, Steven R. and Engel, Jonathan",
    title = "{Double Beta Decay, Majorana Neutrinos, and Neutrino Mass}",
    eprint = "0708.1033",
    archivePrefix = "arXiv",
    primaryClass = "nucl-ex",
    reportNumber = "LA-UR-07-3577",
    doi = "10.1103/RevModPhys.80.481",
    journal = "Rev. Mod. Phys.",
    volume = "80",
    pages = "481--516",
    year = "2008"
}

@article{Engel:2016xgb,
    author = "Engel, Jonathan and Men\'endez, Javier",
    title = "{Status and Future of Nuclear Matrix Elements for Neutrinoless Double-Beta Decay: A Review}",
    eprint = "1610.06548",
    archivePrefix = "arXiv",
    primaryClass = "nucl-th",
    doi = "10.1088/1361-6633/aa5bc5",
    journal = "Rept. Prog. Phys.",
    volume = "80",
    number = "4",
    pages = "046301",
    year = "2017"
}

@article{Albert:2013gpz,
      author         = "Albert, J. B. and others",
      title          = "{Improved measurement of the $2\nu\beta\beta$ half-life
                        of $^{136}$Xe with the EXO-200 detector}",
      collaboration  = "EXO-200",
      journal        = "Phys. Rev.",
      volume         = "C89",
      year           = "2014",
      number         = "1",
      pages          = "015502",
      doi            = "10.1103/PhysRevC.89.015502",
}

@article{KamLAND-Zen:2019imh,
    author = "Gando, A. and others",
    collaboration = "KamLAND-Zen",
    title = "{Precision measurement of the $^{136}$Xe two-neutrino $\beta\beta$ spectrum in KamLAND-Zen and its impact on the quenching of nuclear matrix elements}",
    eprint = "1901.03871",
    archivePrefix = "arXiv",
    primaryClass = "hep-ex",
    reportNumber = "INT-PUB-19-001",
    doi = "10.1103/PhysRevLett.122.192501",
    journal = "Phys. Rev. Lett.",
    volume = "122",
    number = "19",
    pages = "192501",
    year = "2019"
}

@article{Deppisch:2020mxv,
    author = "Deppisch, Frank F. and Graf, Lukas and \v{S}imkovic, Fedor",
    title = "{Searching for New Physics in Two-Neutrino Double Beta Decay}",
    eprint = "2003.11836",
    archivePrefix = "arXiv",
    primaryClass = "hep-ph",
    doi = "10.1103/PhysRevLett.125.171801",
    journal = "Phys. Rev. Lett.",
    volume = "125",
    number = "17",
    pages = "171801",
    year = "2020"
}

@article{Deppisch:2020sqh,
    author = "Deppisch, Frank F. and Graf, Lukas and Rodejohann, Werner and Xu, Xun-Jie",
    title = "{Neutrino Self-Interactions and Double Beta Decay}",
    eprint = "2004.11919",
    archivePrefix = "arXiv",
    primaryClass = "hep-ph",
    doi = "10.1103/PhysRevD.102.051701",
    journal = "Phys. Rev. D",
    volume = "102",
    number = "5",
    pages = "051701",
    year = "2020"
}

@article{Bolton:2020ncv,
    author = "Bolton, Patrick D. and Deppisch, Frank F. and Gr\'af, Luk\'a\v{s} and \v{S}imkovic, Fedor",
    title = "{Two-Neutrino Double Beta Decay with Sterile Neutrinos}",
    eprint = "2011.13387",
    archivePrefix = "arXiv",
    primaryClass = "hep-ph",
    doi = "10.1103/PhysRevD.103.055019",
    journal = "Phys. Rev. D",
    volume = "103",
    number = "5",
    pages = "055019",
    year = "2021"
}

@article{PandaX-4T:2021bab,
    author = "Meng, Yue and others",
    collaboration = "PandaX-4T",
    title = "{Dark Matter Search Results from the PandaX-4T Commissioning Run}",
    eprint = "2107.13438",
    archivePrefix = "arXiv",
    primaryClass = "hep-ex",
    doi = "10.1103/PhysRevLett.127.261802",
    journal = "Phys. Rev. Lett.",
    volume = "127",
    number = "26",
    pages = "261802",
    year = "2021"
}

@article{Chen:2021asx,
    author = "Chen, Xun and others",
    title = "{BambooMC \textemdash{} A Geant4-based simulation program for the PandaX experiments}",
    eprint = "2107.05935",
    archivePrefix = "arXiv",
    primaryClass = "physics.ins-det",
    doi = "10.1088/1748-0221/16/09/T09004",
    journal = "JINST",
    volume = "16",
    number = "09",
    pages = "T09004",
    year = "2021"
}

@article{PANDA-X:2021jua,
    author = "Zhang, Dan and others",
    collaboration = "PANDA-X",
    title = "{Horizontal position reconstruction in PandaX-II}",
    eprint = "2106.08380",
    archivePrefix = "arXiv",
    primaryClass = "physics.ins-det",
    doi = "10.1088/1748-0221/16/11/P11040",
    journal = "JINST",
    volume = "16",
    number = "11",
    pages = "P11040",
    year = "2021"
}

@article{Lenardo_2015,
    author = "Szydagis, M. and others",
    title = "{NEST: A Comprehensive Model for Scintillation Yield in Liquid Xenon}",
    doi = "10.1088/1748-0221/6/10/P10002",
    journal = "JINST",
    volume = "6",
    pages = "P10002",
    year = "2011"
}

@article{Chen:2016qcd,
    author = "Chen, Xun and others",
    title = "{PandaX-III: Searching for neutrinoless double beta decay with high pressure$^{136}$Xe gas time projection chambers}",
    eprint = "1610.08883",
    archivePrefix = "arXiv",
    primaryClass = "physics.ins-det",
    doi = "10.1007/s11433-017-9028-0",
    journal = "Sci. China Phys. Mech. Astron.",
    volume = "60",
    number = "6",
    pages = "061011",
    year = "2017"
}

@article{Ponkratenko:2000um,
    author = "Ponkratenko, O. A. and Tretyak, V. I. and Zdesenko, Yu. G.",
    title = "{The Event generator DECAY4 for simulation of double beta processes and decay of radioactive nuclei}",
    eprint = "nucl-ex/0104018",
    archivePrefix = "arXiv",
    doi = "10.1134/1.855784",
    journal = "Phys. Atom. Nucl.",
    volume = "63",
    pages = "1282--1287",
    year = "2000"
}

@article{Ma:2020kll,
    author = "Ma, Wenbo and others",
    title = "{Internal calibration of the PandaX-II detector with radon gaseous sources}",
    eprint = "2006.09311",
    archivePrefix = "arXiv",
    primaryClass = "physics.ins-det",
    doi = "10.1088/1748-0221/15/12/P12038",
    journal = "JINST",
    volume = "15",
    number = "12",
    pages = "P12038",
    year = "2020"
}

@article{Zhang:2021shp,
    author = "Zhang, D. and others",
    title = "{Rb83/Kr83m production and cross-section measurement with 3.4 MeV and 20 MeV proton beams}",
    eprint = "2102.02490",
    archivePrefix = "arXiv",
    primaryClass = "nucl-ex",
    doi = "10.1103/PhysRevC.105.014604",
    journal = "Phys. Rev. C",
    volume = "105",
    number = "1",
    pages = "014604",
    year = "2022"
}

@article{Haselschwardt:2020iey,
    author = "Haselschwardt, S. J. and Kostensalo, J. and Mougeot, X. and Suhonen, J.",
    title = "{Improved calculations of beta decay backgrounds to new physics in liquid xenon detectors}",
    eprint = "2007.13686",
    archivePrefix = "arXiv",
    primaryClass = "hep-ex",
    doi = "10.1103/PhysRevC.102.065501",
    journal = "Phys. Rev. C",
    volume = "102",
    pages = "065501",
    year = "2020"
}

@article{Qian:2021qiu,
    author = "Qian, Zhicheng and others",
    collaboration = "PandaX-4T",
    title = "{Low radioactive material screening and background control for the PandaX-4T experiment}",
    eprint = "2112.02892",
    archivePrefix = "arXiv",
    primaryClass = "physics.ins-det",
    doi = "10.1007/JHEP06(2022)147",
    journal = "JHEP",
    volume = "06",
    pages = "147",
    year = "2022"
}

@article{Verkerke:2003ir,
    author = "Verkerke, Wouter and Kirkby, David P.",
    editor = "Lyons, L. and Karagoz, Muge",
    title = "{The RooFit toolkit for data modeling}",
    eprint = "physics/0306116",
    archivePrefix = "arXiv",
    reportNumber = "CHEP-2003-MOLT007",
    journal = "eConf",
    volume = "C0303241",
    pages = "MOLT007",
    year = "2003"
}

@article{Rodin:2003eb,
    author = "Rodin, V. A. and Faessler, Amand and Simkovic, F. and Vogel, Petr",
    title = "{On the uncertainty in the 0 nu beta beta decay nuclear matrix elements}",
    eprint = "nucl-th/0305005",
    archivePrefix = "arXiv",
    doi = "10.1103/PhysRevC.68.044302",
    journal = "Phys. Rev. C",
    volume = "68",
    pages = "044302",
    year = "2003"
}

@article{Simkovic:2007vu,
    author = "Simkovic, Fedor and Faessler, Amand and Rodin, Vadim and Vogel, Petr and Engel, Jonathan",
    title = "{Anatomy of nuclear matrix elements for neutrinoless double-beta decay}",
    eprint = "0710.2055",
    archivePrefix = "arXiv",
    primaryClass = "nucl-th",
    doi = "10.1103/PhysRevC.77.045503",
    journal = "Phys. Rev. C",
    volume = "77",
    pages = "045503",
    year = "2008"
}

@article{PandaX-II:2019euf,
    author = "Ni, Kaixiang and others",
    collaboration = "PandaX-II",
    title = "{Searching for neutrino-less double beta decay of $^{136}$Xe with PandaX-II liquid xenon detector}",
    eprint = "1906.11457",
    archivePrefix = "arXiv",
    primaryClass = "hep-ex",
    doi = "10.1088/1674-1137/43/11/113001",
    journal = "Chin. Phys. C",
    volume = "43",
    number = "11",
    pages = "113001",
    year = "2019"
}

@article{XENON:2020iwh,
    author = "Aprile, E. and others",
    collaboration = "XENON",
    title = "{Energy resolution and linearity of XENON1T in the MeV energy range}",
    eprint = "2003.03825",
    archivePrefix = "arXiv",
    primaryClass = "physics.ins-det",
    doi = "10.1140/epjc/s10052-020-8284-0",
    journal = "Eur. Phys. J. C",
    volume = "80",
    number = "8",
    pages = "785",
    year = "2020"
}

@article{Furry:1939qr,
    author = "Furry, W. H.",
    title = "{On transition probabilities in double beta-disintegration}",
    doi = "10.1103/PhysRev.56.1184",
    journal = "Phys. Rev.",
    volume = "56",
    pages = "1184--1193",
    year = "1939"
}

@article{NEXT:2021dqj,
    author = "Novella, P. and others",
    collaboration = "NEXT",
    title = "{Measurement of the ${}^{136}$Xe two-neutrino double beta decay half-life via direct background subtraction in NEXT}",
    eprint = "2111.11091",
    archivePrefix = "arXiv",
    primaryClass = "nucl-ex",
    month = "11",
    year = "2021"
}

@article{NEXT:2015wlq,
    % author = "Mart\'\i{}n-Albo, J. and others",
    author = "Mart\'{i}n-Albo, J. and others",
    collaboration = "NEXT",
    title = "{Sensitivity of NEXT-100 to Neutrinoless Double Beta Decay}",
    eprint = "1511.09246",
    archivePrefix = "arXiv",
    primaryClass = "physics.ins-det",
    reportNumber = "FERMILAB-PUB-16-669-CD-ND",
    doi = "10.1007/JHEP05(2016)159",
    journal = "JHEP",
    volume = "05",
    pages = "159",
    year = "2016"
}

@article{EXO-200:2019rkq,
    author = "Anton, G. and others",
    collaboration = "EXO-200",
    title = "{Search for Neutrinoless Double-$\beta$ Decay with the Complete EXO-200 Dataset}",
    eprint = "1906.02723",
    archivePrefix = "arXiv",
    primaryClass = "hep-ex",
    doi = "10.1103/PhysRevLett.123.161802",
    journal = "Phys. Rev. Lett.",
    volume = "123",
    number = "16",
    pages = "161802",
    year = "2019"
}

@article{KamLAND-Zen:2016pfg,
    author = "Gando, A. and others",
    collaboration = "KamLAND-Zen",
    title = "{Search for Majorana Neutrinos near the Inverted Mass Hierarchy Region with KamLAND-Zen}",
    eprint = "1605.02889",
    archivePrefix = "arXiv",
    primaryClass = "hep-ex",
    doi = "10.1103/PhysRevLett.117.082503",
    journal = "Phys. Rev. Lett.",
    volume = "117",
    number = "8",
    pages = "082503",
    year = "2016",
    note = "[Addendum: Phys.Rev.Lett. 117, 109903 (2016)]"
}

@article{Agostini:2017jim,
    author = "Agostini, Matteo and Benato, Giovanni and Detwiler, Jason",
    title = "{Discovery probability of next-generation neutrinoless double- \ensuremath{\beta} decay experiments}",
    eprint = "1705.02996",
    archivePrefix = "arXiv",
    primaryClass = "hep-ex",
    doi = "10.1103/PhysRevD.96.053001",
    journal = "Phys. Rev. D",
    volume = "96",
    number = "5",
    pages = "053001",
    year = "2017"
}

@article{PandaX-II:2021jmq,
    author = "Yan, Binbin and others",
    collaboration = "PandaX-II",
    title = "{Determination of responses of liquid xenon to low energy electron and nuclear recoils using a PandaX-II detector}",
    eprint = "2102.09158",
    archivePrefix = "arXiv",
    primaryClass = "physics.ins-det",
    doi = "10.1088/1674-1137/abf6c2",
    journal = "Chin. Phys. C",
    volume = "45",
    number = "7",
    pages = "075001",
    year = "2021"
}

@article{Zhu:2021qss,
    author = "Zhu, Shaofei and McCutchan, E. A.",
    title = "{Nuclear Data Sheets for A=214}",
    doi = "10.1016/j.nds.2021.06.001",
    journal = "Nucl. Data Sheets",
    volume = "175",
    pages = "1--149",
    year = "2021"
}

@article{PandaX:2018wtu,
    author = "Zhang, Hongguang and others",
    collaboration = "PandaX",
    title = "{Dark matter direct search sensitivity of the PandaX-4T experiment}",
    eprint = "1806.02229",
    archivePrefix = "arXiv",
    primaryClass = "physics.ins-det",
    doi = "10.1007/s11433-018-9259-0",
    journal = "Sci. China Phys. Mech. Astron.",
    volume = "62",
    number = "3",
    pages = "31011",
    year = "2019"
}

@article{Cleveland:1979,
    author = {Cleveland, William S.},
    title = {Robust Locally Weighted Regression and Smoothing Scatterplots},
    journal = {Journal of the American Statistical Association},
    volume = {74},
    number = {368},
    pages = {829-836},
    year  = {1979},
    doi = {10.1080/01621459.1979.10481038},
    URL = {https://www.tandfonline.com/doi/abs/10.1080/01621459.1979.10481038},
}

@article{Dolinski:2019nrj,
    author = "Dolinski, Michelle J. and Poon, Alan W. P. and Rodejohann, Werner",
    title = "{Neutrinoless Double-Beta Decay: Status and Prospects}",
    eprint = "1902.04097",
    archivePrefix = "arXiv",
    primaryClass = "nucl-ex",
    doi = "10.1146/annurev-nucl-101918-023407",
    journal = "Ann. Rev. Nucl. Part. Sci.",
    volume = "69",
    pages = "219--251",
    year = "2019"
}

@article{Cremonesi:2013vla,
    author = "Cremonesi, O. and Pavan, M.",
    title = "{Challenges in Double Beta Decay}",
    eprint = "1310.4692",
    archivePrefix = "arXiv",
    primaryClass = "physics.ins-det",
    doi = "10.1155/2014/951432",
    journal = "Adv. High Energy Phys.",
    volume = "2014",
    pages = "951432",
    year = "2014"
}

@article{Aprile:2022qou,
    author = "Aprile, Elena and others",
    title = "{Double-Weak Decays of $^{124}$Xe and $^{136}$Xe in the XENON1T and XENONnT Experiments}",
    eprint = "2205.04158",
    archivePrefix = "arXiv",
    primaryClass = "hep-ex",
    month = "5",
    year = "2022"
}

@article{Yang:2021hnn,
    author = "Yang, Jijun and others",
    title = "{Readout electronics and data acquisition system of PandaX-4T experiment}",
    eprint = "2108.03433",
    archivePrefix = "arXiv",
    primaryClass = "physics.ins-det",
    doi = "10.1088/1748-0221/17/02/T02004",
    journal = "JINST",
    volume = "17",
    number = "02",
    pages = "T02004",
    year = "2022"
}

@misc{Kotila_SSD,
  author = "J. Kotila",
  year = "2022",
  howpublished = "Private communication"
}

@article{Goeppert-Mayer:1935uil,
    author = "Goeppert-Mayer, M.",
    title = "{Double beta-disintegration}",
    doi = "10.1103/PhysRev.48.512",
    journal = "Phys. Rev.",
    volume = "48",
    pages = "512--516",
    year = "1935"
}

@article{Agostini:2020cpz,
    author = "Agostini, Matteo and Bossio, Elisabetta and Ibarra, Alejandro and Marcano, Xabier",
    title = "{Search for Light Exotic Fermions in Double-Beta Decays}",
    eprint = "2012.09281",
    archivePrefix = "arXiv",
    primaryClass = "hep-ph",
    reportNumber = "TUM-HEP 1306/20",
    doi = "10.1016/j.physletb.2021.136127",
    journal = "Phys. Lett. B",
    volume = "815",
    pages = "136127",
    year = "2021"
}

@Misc{RGA,
howpublished = {\url{https://www.hidenanalytical.com/products/residual-gas-analysis/}},
}

@Misc{NIST,
howpublished = {\url{https://webbook.nist.gov/}},
}
\end{document}